# A new watermarking method to protect blockchain records comprising handwritten files

1st Rustam Latypov
Institute of computational mathematics
Kazan federal university
Kazan, Russia
Roustam.Latypov@kpfu.ru

2nd Evgeni Stolov
Institute of computational mathematics
Kazan federal university
Kazan, Russia
ystolov@kpfu.ru

*Abstract*—**A new type of watermarks for handwritten black-white documents is suggested. Insertion of the watermark in a document minimizes distortion of the latter. The method is intended for validation of handwritten records placed in blockchain database.**
*Index Terms*—**blockchain, watermark, handwritten document**

## I. INTRODUCTION

Blockchain is one of the most promising recent technologies, and many authors denote pile of the use cases like the land registry, real estate, voting, recordkeeping, etc. [1]–[3]. A regular blockchain database contains records of the same type. Let us say, each record is a set of personal data of a human being. It may be that a black-white handwritten document is presented in any record. For example, it is the case when the results of a written test must be saved.

The blockchain usually stores information about a transaction and one need a way to ensure some certainty about digital files that are involved in those transactions. The simplest type of this is to use a unique content file identifier in the transaction stored in the blockchain as well as insert that same identifier in the content file. There are two possible solutions of the problem: a hash code followed each document in the record or a watermark is inserted in each document [4]–[6]. Any of the approaches have its advantages and drawbacks. Usage of a hash key increases the size of the database but the hash calculation is simple. If somebody reveals the presence of a hash, he/she will try to break the protection. Implementation of a watermark does not add any extra information in the record, but inserting a watermark into the handwritten document and its extraction takes a time. On the other hand, since the watermark is hidden there is no sense for looking at how to avoid the security.

The problem of upcoming handwriting databases has received sizable attention in recent years (for example [7], [8]). These databases can be employed in result comparison of new methods related to handwriting analysis. Suppose that after a routine observation, one detects the following error in the database: for some two sequential records, the hash of the first record does not match the value stored in the second record. It means that the database contains a corrupted record. The corrupted record can be revealed effortlessly by means of regular software included in the package. The other problem is to find out which part of the documents in the record was changed. In the case of the public blockchain, the corrupted database can be compared with a clean copy, though the compare takes some resources. In the case of the private database, one has to keep a copy of the base. It means that all the advantages of implementation of the blockchain technology will be lost [9].

In this paper, we propose a technique which involves two parts of the same record. Each of two parts of the record is leveraged as a hash of the other part. The case, where any record of the database contains a handwritten black-white document and more text information, is investigated. That text information is used as a watermark and it is inserted into the handwritten black-white document. If a watermark is not detected in the handwritten document that document was changed. If the watermark exists and does not coincide with the text information the latter was changed. We faced the problem while the results of a written examination in math must be saved. In this paper, we will illustrate the developed method by the example of written test folios of students.

Our paper is organized as follows. In Section II, we describe the problem of watermarking in handwritten papers intended for inclusion in the blockchain. We plan the idea for our algorithm in Section III. Section IV describes the algorithms for watermark inclusion and extraction for handwritten text. The problem of size selection for pieces of manuscript those picked up for watermarking is discussed in Section V. In VI we talk over the security issues in implementing our algorithm, after that we make conclusions.

## II. DESCRIPTION OF THE PROBLEM

Before the exam, each student gets an individual task and he/she must present a solution on a white sheet of paper. The folio was scanned and converted into a black-white document using Otsu algorithm [10]. Two examples of pages with solutions filled in during the exam in linear algebra are presented in Fig. 1 and Fig 2.

Each examination page is a part of a record, that is placed into the standard blockchain database. The second part of the record is additional text (metadata) information related to the exam: the date, the name of the student, the name of the course, the group, and the index of the task. An illegal change of the page to another page in the record or correction the additional information is considered as a possible attack to the database. It is known [2] that the last record in blockchain database is not protected until the next record is added. The direct transform of the blockchain file, although such intrusion can be revealed by the evident procedure, is also considered as a possibility.

Fig. 1. Example 1 of examination page.

Fig. 2. Example 2 of examination page

Our idea to protect the record is as follows: watermark is taken equal to the textual information (metadata) which follows the handwritten document. We insert the watermark in the handwritten document. This way we combine both methods - watermark is inserted into the picture and additional information becomes a hash of the picture.

There are well-known techniques for placing watermark into the printed black-white text document: change spaces between words or lines, small changes of a grapheme, and others. The watermarks, based upon those properties, can be extracted since there are standard for all of the mentioned features. If one deals with a black-white handwritten document, neither of the methods can be implemented. There are also no pictures which are used for placing various types of watermark [11]. In this paper, we suggest a new type of watermarks that use special features of the container. With the given watermark in the picture, all changes of the additional information will be detected.

III. IDEA FOR WATERMARKING

The idea of the suggested watermark is as follows. Let us select from the page a horizontal chunk, having the height $Step$ and starting from the beginning of the page. The reasons for choosing the value of $Step$ will be discussed later. Regular black-white handwritten text can be thought of as a set of ones and zeros. We use 1's for designation black pixels.

Let us introduce some notations. Let the beforehand given watermark for insertion is presented as a string $Str$ of bits, and let $StrLn$ be the length of $Str$. Let $Num(L)$ be the number of 1's on the vertical line $L(x)$ drawn inside the chunk at distance $x$ from the left bound of the page. Let us set a correspondence between vertical lines $L(x)$ with $Num(L) \neq 0$ and bits $Bit \in Str$. Suppose that bit $Bit$ relates to the line $L(x)$. If $Bit = 1$ and $Num(L)$ is odd or $Bit = 0$ and $Num(L)$ is even, then $L(x)$ is stayed unchanged otherwise we add a black pixel (if $Bit = 0$) to the line or delete such pixel ($Bit = 1$). As a result, the parity of the number of 1's on the line $L(x)$ coincides with the bit $Bit$ corresponding to the line. If all the available vertical lines from the chunk are exhausted but the number of the bits in $Str$, that are coded by the lines, is less than $StrLn$, a new chunk of the same height $Step$ is selected from the rest of the page, and the procedure is continued.

For extraction of the watermark, one has to select the chunks from the modified page, to count *modulo 2* of 1's on each vertical line that was used for coding and restore the bits of the watermark.

## IV. Algorithm

In this section, we present some details related to the realization of the idea described above. In the previous section, it was mentioned that not all vertical lines are good for coding. In what follows, an exact definition of the available lines will be given.

According to the idea described in the previous section, the page is divided into the strips of the height $Step$. Let $L$ be a vertical line in the strip and a function $Fit$ that return values $True$ or $False$ depending on the fitness of $L$ for coding. Let $L'$ be a result of the transformation of $L$ while coding. Correct decoding procedure is possible only if
$$Fit(L) = Fit(L') \quad (1)$$
is fulfilled. The procedure modifying a vertical line for coding the current bit must not lead to the appearance of new black pixels outside of the strip under consideration. If a pixel is set outside the strip, it will be excluded while extraction of the watermark by means of the chunk of the same height, thus the watermark will be restored incorrectly. Let us set the following rules:

1) $Fit(L) = True\ iff\ Num(L) >= Strip/2$.
2) If $Fit(L) = True$ And $Num(L) < .2\ Strip/3$ one adds pixel to $L$ otherwise deletes pixel during transformation.
3) If a pixel is added to $L$ the new line $L'$ has no pixels outside of the current strip.

When someone deletes a pixel from a line, no new pixel appears outside of the strip. The problem is to meet this condition while inserting a new pixel in the line. Later we will show the way our algorithm guarantees that property.

All the distortions of the text must be imperceptible for the reader. The symbol in the text, overlapping with the modified line, has to keep its sense. To meet these conditions, one firstly investigates whether gaps exist in the line while adding a new black pixel to the line (see Fig. 3). If there is a gap, one tries to place the new pixel inside of the gap with maximal length. If $Num(L) > 2\ Step/3$ and the line has no gaps the first black pixel is deleted. The procedure described above is realized by Algorithm 1. There is the function $Flag, Pos = getPosition(L)$ that performs all "intellectual" work. If $Flag = 1$ then $L$ is not good for the modification, otherwise, the line can be altered. If $Flag = 0$ the number of ones in $L$ is even, and if $Flag = 1$ then this value is odd. $Pos$ is the position in $L$ which must be changed. The function $changeLine(L. Pos)$ implements required transformation of $L$ by inserting to $L$ or removing from $L$ pixel at position $Pos$ in line depending upon value $Num(L)$. The Algorithm 2 presents all details of realization $getPosition(L)$. Here is supposed that the watermark can be placed into a single strip. Spreading the algorithm to the case, where a few strips are used for inserting the watermark, is evident. Extraction of the watermark is performed by Algorithm 3.

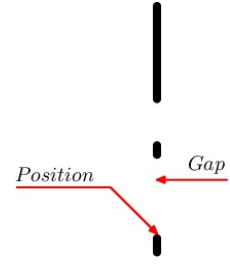

Fig. 3. Possible position for insertion extra black pixel in line with gaps

**Algorithm 1.** Coding bits by vertical lines in chunk. Watermark is placed into columns of $Matr$

1: $Chunk \leftarrow Picture, Step$ {Select chunk}
2: $Matr \leftarrow Chunk$ {$Matr$ consits of 0's and 1's}
3: $Col \leftarrow Matr$ {number of columns}
4: $Str \leftarrow Watermark$ {watermark as a string of bits}
5: $StrLn \leftarrow Str$ {length of watermark}
6: $Count \leftarrow 0$
7: **for** $x = 0\ to\ Col - 1$ **do**
8:   **if** $Count = StrLn$ **then**
9:     *break*
10:   **end if**
11:   $L \leftarrow Matr[:,x]$ {column of matrix}
12:   $Flag, Pos \leftarrow getPosition(L)$   {$Flag \in \{1, 0, 1\}$}
13:   **if** $Flag = -1$ **then**
14:     *continue*
15:   **else**
16:     $Bit \leftarrow Str[Count]$
17:     $Count \leftarrow Count + 1$
18:     **if** $Bit = Flag$ **then**
19:       *continue*
20:     **else**
21:       $changeLine(L.Pos)$ {changing parity of the number of black pixels on $L$}
22:     **end if**
23:   **end if**
24: **end for**

## V. Choice the height of chunk

Let us discuss the problem of choosing the height of the strip (parameter $Step$ in the algorithms). This value must be the same for pictures in all records since the coding procedure is independent of the picture. In practice, it is impossible to create pictures of the same size. The experiments show that all watermarks after bit encoding have a length no more than 1300 bits in our case, so the choice of the height must provide a possibility for placement watermark in the strips taken from one page. That is the reason, one has to evaluate the number of lines available for coding. Comparing Fig 1 and Fig. 2, one can see that there is a substantial difference in textures of those pages. Let *D* be the number of bits coded by means of one page.

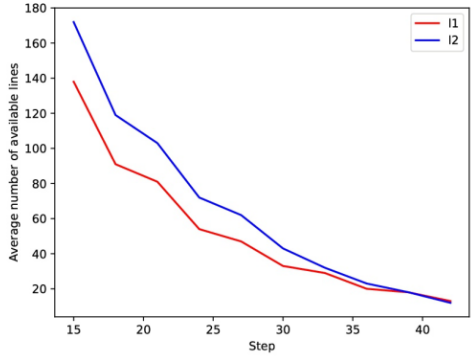

Fig. 4. Dependence of $D$ upon the height of chunk. Label 'i1' relates to Fig. 1 and label 'i2' - to Fig. 2

The number $D$ is leveraged as a basic parameter while choosing the height of the chunk. Of course, the result depends on the type of the text in the picture. A comparison of the values of the parameter $D$ for both the examples "Example 1" and "Example 2" is given on the Fig. 4.

On the base of that graphs, we propose an empirical rule for choice of the height of chunk in the text. It follows from Fig. 4 that the value 40 pixels leads to equal values of $D$ for both the examples. Our experiments show that this value provides acceptable results for all considered examples. It may be interesting to note, that this value is close to the average height of symbols in all documents. It must be mentioned that according to Fig. 4 just a few columns in the strip are good for placement of a bit of watermark (total numbers of columns in the picture equals 1763).

**Algorithm 2.** Function *getPorition(L)*
1: $LArray \leftarrow L$ { convert $L$ into 1D array }
2: $LnL \leftarrow LArray$ { length of array }
3: $Num \leftarrow LArray$ { sum of black pixels }
4: **if** ($Num < LnL/2$) **then**
5:     **return** $(-1, -1)$ { the line is not good for coding}
6: **end if**
7: $Num2 \leftarrow Num$ { sum of pixels modulo 2}
8: **if** ($Num > 2LnL/3$) **then**
9:   **for** $x = 0$ to $LnL - 1$ **do**
10:     **if** $LArray[x] = 1$ **then**
11:       **return** $(Num2, x)$ {the line fits, delete pixel}
12:     **end if**
13:   **end for**
14:   **else**
15: $GapLst$ $LArray$ {form list of bounds of gaps in array}
16: **if** $length(GapList) = 0$ **then**
17:   **for** $x = 0$ to $LnL - 1$ **do**
18:     **if** $LArray[x] = 0$ **then**
19:       **return** $(Num2, x)$
23: $(A, B) \leftarrow (GapList)$ {bounds of the gap of maximal length from the list}
24:     **return** $(Sm2, B)$
25:   **end if**
26: **end if**

**Algorithm 3** Extraction of watermark. All extracted bits are placed into $ListOfBits$
1: $Matr \leftarrow Chunk$ {$Matr$ consits of 0's and 1's}
2: $Col \leftarrow Matr$ {number of columns}
3: $Str \leftarrow Watermark$ {watermark as a string of bits}
4: $StrLn \leftarrow Str$ {length of watermark}
5: $Count \leftarrow 0$
6: $ListOfBits \leftarrow Empty$ {all extracted bits are collected here}
7: **for** $x = 0$ to $Col - 1$ **do**
8:   **if** $Count = WtrLn$ **then**
9:     **if** $Str = ListOfBits$ **then**
10:       **return** true
11:     **else**
12:       **return** false
13:     **end if**
14:   **end if**
15:   $L \leftarrow Matr[:,x]$ {column of matrix}
16:   $Flag, Pos \leftarrow getPosition(L)$ { $Flag \in \{1, 0, 1\}$}
17:   **if** $Flag = -1$ **then**
18:     *continue*
19:   **else**
20:     $ListOfBits.append(Flag)$
21:     $Count \leftarrow Count + 1$
22:   **end if**
23: **end for**
24: **return** false

## VI. NOTES ABOUT CRYPTOGRAPHIC SECURITY

The result of insertion watermark into the black-white picture by the described algorithms cannot be discovered via immediate observation. On the other hand, the procedure is not a very reliable one from point of view of cryptographic security. It depends on the single key — the parameter $Step$. Nevertheless, let us compare values of two autocorrelation functions for chunks before and after inserting a watermark. That is the first step when somebody is trying to reveal the presence of the watermark. The function $ACorr$ is autocorrelation function of the array containing the parity of numbers of black pixels in vertical lines from the chunk. While calculation, all the columns with non-zero numbers of pixels are considered whereas the watermark was inserted only into available vertical lines. The phrase in Russian (Fig. 5) was converted into a bit sequence after $utf8$ encoding and was leveraged as a watermark. The length of the bit sequence equals 909.

20.06.2018, К.Иванов, группа 941, Линейная алгебра и геометрия, задание №4

Fig. 5. Phrase as a watermark

The correlation function before inserting water-mark is presented in Fig. 6.

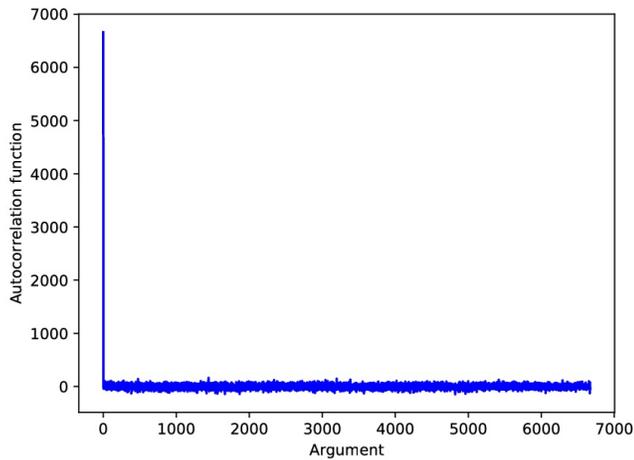

Fig. 6. Values of autocorrelation function before inserting watermark

There is a natural maximum of $ACorr$ at 0 that is 6661 ( the values of $ACorr$ are not normalized). But it is interesting to observe the values $ACorr$ for arguments from the interval [1,5]. Those are 273, 15, 7, 40, 61. The value $ACorr[1]$ exceeds the values $ACorr[x]$, $x > 1$. This phenomenon re- flects the situation where the adjacent vertical lines from chunk coincide if the width of the curves takes more than one pixel. In Fig. 7, differences of the autocorrelation functions before and after insertion of the watermark for various values of $Step$ are presented. It is easy to see that all the differences have a random form, so investigation of autocorrelation functions cannot indicate the presence of the watermark in the picture.

## VII. CONCLUSION

This paper introduces a method to protect records containing black-white images and text information in the database. One transforms the text information into a bit string, which is placed in the image as a watermark. The results of the insertion are invisible by a person. The appropriate algorithm uses only simple operations. Interposed watermark is secure against attacks that are based on statistical analysis. The method can apply for protection blockchain databases

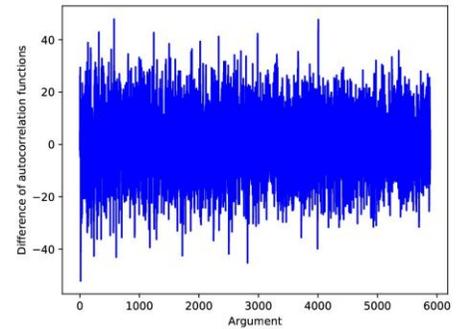

(a) Step=30

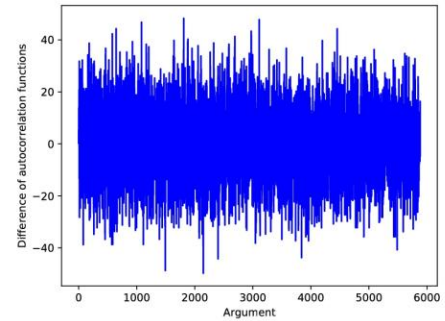

(b) Step=40

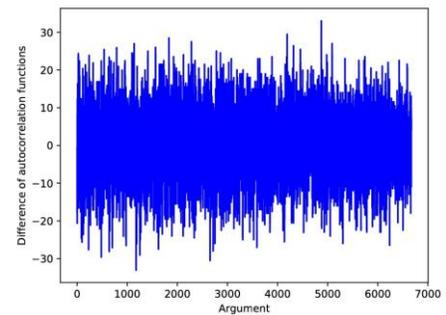

(c) Step=50

Fig. 7. Differences of autocorrelation functions before and after inserting watermark

## REFERENCES

[1] J. Umeh,"Beyond Bitcoin and the Blockchain," ITNOW, 60(3), pp.48-49, 2018.
[2] V.Lemieux, "A typology of blockchain recordkeeping solutions and some reflections on their implications for the future of archival preserva- tion," Proc. IEEE Int. Conf. on Big Data (Big Data), pp.2271–2278,2017.
[3] A.Galiev, S.Ishmukhametov, R.Latypov, N.Prokopiev, E.Stolov, I.Vlasov, "ARCHAIN: A Novel Blockchain Based Archival System," Proc. IEEE Second World Conference on Smart Trends in Systems, Security and Sustainability, 2018 (in press)
[4] A. Swaminathan, Y. Mao, M. Wu, "Robust and secure image hashing," IEEE Trans. on Information Forensics and Security, 1(2),pp.215–230, 2006.
[5] Z.Meng, T. Morizumi, S. Miyata, H. Kinoshita, "Design Scheme of Copyright Management System Based on Digital Watermarking and Blockchain," Proc. IEEE 42nd Annual Computer Software and Appli- cations Conf. (COMPSAC), 2, pp.359–364, 2018.
[6] R. Latypov, E.Stolov, "A New Method for Slant Calculation in Off-Line Handwriting Analysis," Proc.IEEE 41st Int. Conf. on Telecommunica- tions and Signal Processing (TSP), pp.1–5,2018.
[7] C. Liu, F. Yin, D. Wang, Q. Wang, "CASIA Online and Offline Chinese Handwriting Databases," Proc. IEEE Int. Conf on Document Analysis and Recognition, pp.37–41, 2011.
[8] C. Djeddi, A.Gattal, L. Souici-Meslati, I.Siddiqi, Y. Chibani, H. El Abed, "LAMIS-MSHD: A Multi-script Offline Handwriting Database,"Proc. 14th Int. Conf. on Frontiers in Handwriting Recognition, pp.93–97, 2014.
[9] D. Guegan. "Public Blockchain versus Private blockchain." Documents de travail du Centre dEconomie de la Sorbonne 2017.20 - ISSN : 1955-611X. 2017.
[10] N. Otsu, "A threshold selection method from gray-level histograms," IEEE Trans. Sys., Man., Cyber. 9, pp. 62–66, 1979.
[11] Peter Wayner, Disappearing Cryptography: Information Hiding: Steganography and Watermarking. 3rd ed., Elsevier Inc, 2009.